\documentclass[11pt]{iopart}
\usepackage{iopams,setstack}

\newcommand{\dif}{\,\mathrm{d}}		  % Define dif

\newcommand{\diag}{\mathrm{diag}}         % Define diag operator
\usepackage{psfrag,color}

\usepackage[hyperfootnotes=false]{hyperref}
\hypersetup{
 colorlinks=true,
 citecolor=blue,
 urlcolor=blue,
 allcolors=blue}
\usepackage{enumerate,cite,latexsym,graphicx}

\newtheorem{theorem}{Theorem}

\newtheorem{definition}{Definition}

\begin{document}

\title[]{Design of coherent quantum observers for linear quantum systems}
\author{Shanon L Vuglar\textsuperscript{1} and Hadis Amini\textsuperscript{2}}%

\address{\textsuperscript{1}
School of Information Technology and Electrical Engineering, 
University of New South Wales at the Australian Defence Force Academy, Canberra ACT 2600, Australia
}
\ead{\color{blue}shanonvuglar@vuglar.com}
\address{\textsuperscript{2}
	Edward L. Ginzton Laboratory, Stanford University, Stanford, CA 94305, USA}
\ead{\color{blue} nhamini@stanford.edu}
%\address{\textsuperscript{3}
	%Edward L. Ginzton Laboratory, Stanford University, Stanford, CA 94305, USA}
%\ead{\color{blue}hmabuchi@stanford.edu}
\begin{abstract}
Quantum versions of control problems are often more difficult than their classical 
counterparts because of the additional constraints imposed by quantum dynamics. 
For example, the quantum LQG and quantum $H^\infty$ optimal control problems remain open.
To make further progress, new, systematic and tractable 
methods need to be developed. This paper gives three algorithms for  
designing coherent quantum observers, i.e., quantum 
systems that are connected to a quantum plant and their outputs provide information about 
the internal state of the plant. Importantly, coherent quantum observers avoid measurements of 
the plant outputs. We compare our coherent quantum observers with a classical (measurement-based) 
observer by way of an example involving an optical cavity with thermal and vacuum noises as inputs. 
\end{abstract}
%\pacs{1315, 9440T}
%\submitto{\JPG}
\maketitle
 \tableofcontents
%\addcontentsline{toc}

%%%%%%%%%%%%%%%%%%%%%%%%%%%%%%%%%%%%%%%%%%%%%%%%%%%%%%%%%%%%%%%%%%%%%%%%%%%%%%%%
\section{Introduction} \label{sec:intro}
Feedback control of quantum systems can be broadly categorized into two schemes: 
`classical' (or measurement-based) and `coherent' control. Classical control 
involves making measurements on the plant (for example homodyne 
or heterodyne detection in the case of optical systems) and then generating feedback 
control signals based on these measurements. For a treatment of this topic see 
for example \cite{WM10}. In this paper, we are concerned with coherent control which 
uses controllers that are themselves quantum systems, coupled directly to the plant. 
One advantage of coherent control schemes is that they 
avoid the loss of quantum information that occurs 
during measurements. Coherent quantum control is an active research area 
\cite{JNP1,NJP1,MaP4,Vladimirov:2013et,MAB08,JG08,ZJ11,ISY11,Hamerly:2012wi}
and recent results  \cite{Hamerly:2012wi} indicate regimes in which coherent 
controllers perform better than the optimal classical controllers. Coherent quantum observers represent an important building block in developing systematic 
and tractable approaches to coherent control problems.

Despite recent progress, quantum versions of the 
$H^\infty$ \cite{JNP1,MaP4} and LQG \cite{NJP1,Vladimirov:2013et} optimal control 
problems remain open. These problems are difficult because of the constraints 
on the class of allowable controllers: quantum systems must evolve unitarily and preserve 
commutation relations \cite{AFP09,Wein13}. These constraints lead to the notion of 
\emph{physical realizability} \cite{JNP1,VuP11a,VuP12a,VuP12c}. Current approaches 
\cite{JNP1,MaP4,NJP1,Vladimirov:2013et} are 
only tractable for relatively simple examples.

Feedback control schemes require that the controller has access to information about 
the internal state of the plant. For example, in the classical LQG problem, 
%(see for example \cite{KS72}), 
the Kalman filter \cite{KS72,speyer2008stochastic} is used to obtain an optimal estimate 
of the plant's internal state from a series of noisy measurements. In coherent control 
schemes, the controller does not have access to measurements, rather it must make use 
of information from its direct coupling with the plant. 
\begin{figure}[h]
		\includegraphics[trim = 0mm 0mm 0mm 0mm, scale=0.8]{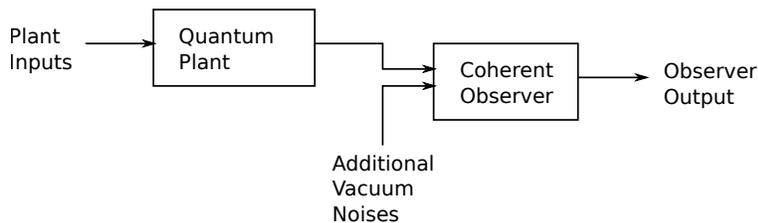}
		\caption{Quantum plant and coherent quantum observer.}
		\label{fig:intro}	
\end{figure}
A coherent quantum observer is a quantum system that is designed such that when directly coupled 
with a plant, its outputs provide information about the internal state of the plant.
The coherent quantum observer can in turn be directly coupled to control inputs of the plant, 
possibly via intermediate quantum systems, to achieve desired control outcomes whilst avoiding 
measurement.

Different approaches to design coherent quantum observers are discussed in 
\cite{JNP1,Miao:2012ug,Vladimirov:2013ti}.
However, as with the quantum control problems mentioned above, the main difficulties 
come from the fact that a coherent quantum observer should satisfy 
\emph{physical realizability} constraints. 

In this paper, we extend previous \emph{physical realizability} results 
\cite{VuP11a,VuP12a,VuP12c} to obtain algorithms for designing coherent quantum observers. 
The previous results we use demonstrate how strictly proper, 
linear time invariant (LTI) systems can be made physically realizable by allowing
additional quantum noises. 
We apply these results to construct coherent quantum observers using a Kalman filter 
which is modified by adding quantum noises as prescribed in \cite{VuP11a,VuP12a,VuP12c}.
%
%New for final version
%
%This is similar to the approach taken in \cite{JNP1}, the main difference being that 
%in our approach only necessary quantum noises are added which can lead to a better performance. However, these approaches, while tractable, 
%are generally suboptimal.
%
This is different to the approach taken in \cite{JNP1} because we make use of the stronger 
results from \cite{VuP11a,VuP12a,VuP12c} so that only necessary quantum noises are added. 
We also incorporate novel refinements not considered in 
\cite{JNP1,Miao:2012ug,Vladimirov:2013ti}. 
Our approach is tractable and can lead to better performance, however the 
coherent observers obtained are generally suboptimal.
%
%end new for final version

We now outline three algorithms that we propose. 
The first algorithm is based on the Kalman filter which is modified by allowing additional 
vacuum noise inputs such that the resulting system is physically realizable. 
%
% Removed for final version:   ... which gives a similar method considered in~\cite{JNP1}.
%
The second 
algorithm attempts to improve on the first by incorporating a free parameter over which we 
optimize. The purpose of this parameter is to compensate for the effect of the 
additional quantum vacuum noises. The third algorithm attempts to find a state transformation 
of the Kalman filter such that it can be made physically realizable with the minimal 
number of additional quantum noises. 
Despite being suboptimal estimations, these algorithms provide a systematic and tractable approach to 
coherent quantum observer design.
Like the celebrated Kalman filter, it is envisaged that the coherent quantum observer will play an 
important role in the solution to coherent control problems.

The main contribution of this paper is to give
two additional algorithms for coherent quantum observer 
design which incorporate novel refinements not considered in 
\cite{JNP1,Miao:2012ug,Vladimirov:2013ti}.
We compare the performance of our three different algorithms 
using a metric corresponding to the steady-state expected value of the 
symmetrized error covariance matrix and show that the novel refinements which we propose 
can lead to better perfomance than implementing a 
Kalman filter as a quantum system as in our first algorithm.

The remainder of this paper is structured as follows. In Section \ref{sec:models}, we introduce  
models to describe the quantum systems given by linear quantum stochastic differential equations. In Section \ref{sec:pr}, we formally define 
\emph{Physical Realizability} of such linear quantum systems. 
We are then able to give our problem formulation in Section \ref{sec:formulation} where we also present relevant existing results which we will utilize. 
Section \ref{sec:observers} contains the main contribution of the paper, we present three algorithms 
for designing coherent quantum observers for linear quantum systems. In Section~\ref{sec:classical}, we present a measurement-based (classical) observer. The three algorithms are then compared with each other and the 
measurement-based alternative by way of an example in Section \ref{sec:example}. Finally, we give our 
conclusion in Section \ref{sec:conclusion}.
\section{Linear quantum system models} \label{sec:models}
Consider a quantum system defined on a Hilbert space $\mathcal H$, and its environment modeled by the bosonic or symmetric Fock space over the Hilbert space $L^2(\mathbb R_+)$ of square integrable wave functions on the real positive line, corresponding to a single boson field mode. The evolution of the 
composite system, which is a closed system, can be described by a unitary operator~$U$ acting on the 
tensor product $\mathcal H\otimes \mathcal F$ that obeys the following quantum stochastic 
differential equations (QSDEs) as described by  Hudson and Parthasarathy~\cite{HP84} 
\begin{equation*}
dU(t)=\left(db^\dag L-L^\dag db-\frac{1}{2} L^\dag L\,dt-iH\,dt\right)U(t),\quad U(0)=I.
\end{equation*}
Here, $H$ corresponds to the Hamiltonian of the system, $L$ describes the coupling between the system 
and the environment, 
%new for final version
and ${X}^\dagger$ denotes the adjoint of an operator $X$. 
%end new
The operators $b$ and $b^\dag$ are the annihilation and creation processes defined 
on $\mathcal F.$ 

In the Heisenberg picture, the evolution of a self-adjoint operator $x$ is described by 
\begin{equation}\label{eq:evox}
x(t)=U(t)^\dag(x(0)\otimes I)U(t).
\end{equation}
Using the input-output formalism of \cite{gardiner1985input}, we also have 
\begin{equation}~\label{eq:evoy}
y(t)=U(t)^\dag(I \otimes w(t))U(t)
\end{equation}
where $y(t)$ is the output of the system and $w(t)$ is its input.
Here, the self-adjoint entries of the vector $w(t)$ which act on the 
Boson Fock space $\mathcal{F}$
correspond to the quantum noises driving the system \cite{HP84}. 
The noise increments $dw(t)$ in quadrature form are given by
\begin{equation}~\label{eq:wiener}
dw=
	\left[ \begin{array}{cc}
		db(t)+db(t)^\dag \\
		i(db(t)^\dag-db(t))
	\end{array} \right].
	\end{equation}
	Generally speaking, the QSDEs for a given quantum system can be obtained by applying quantum It\={o} rules to $x$ and $y$ which satisfy dynamics~(\ref{eq:evox}) and~(\ref{eq:evoy}) respectively, and using the following quantum It\={o}  multiplication table~\cite{HP84,parthasarathy2012introduction}:
\begin{equation}~\label{eq:table}
db\,db=0, \quad db\,db^\dag=(1+k_n)dt, \quad db^\dag\, db=k_ndt, \quad {\rm and} \quad db^\dag \,db^\dag=0.
\end{equation}
Also, by using $(dt)^2=0$, and $dtdb=0=dtdb^\dag.$
Here, $k_n$ is a parameter describing the intensity of the thermal noise input. The special case where 
$k_n = 0$ corresponds to an input being a vacuum noise.

\medskip

In the case of open quantum harmonic oscillators, which we consider in this paper, the Hamiltonian $H$ 
is quadratic and the coupling operator $L$ is linear. This leads to linear QSDEs of the form
\begin{eqnarray}
	\dif x(t) &=& A x(t) \dif t + B \dif w(t),\nonumber \\
	\dif y(t) &=& C x(t) \dif t + D \dif w(t),
	\label{eqn:plant}
\end{eqnarray}
where $A,$ $B,$ $C,$ and $D$ are real matrices which are supposed to be in $\mathbb R^{n_x\times n_x},$ $\mathbb R^{n_x\times n_w},$ $\mathbb R^{n_y\times n_x}$ and $\mathbb R^{n_y\times n_w}$ respectively, and $n_x,$ $n_w,$ and $n_y$ are positive integers. However, not all QSDEs of the form (\ref{eqn:plant}) correspond to open quantum harmonic oscillators. 
When they do, they are said to be physically realizable. This is explained further in the following section 
where we also give explicit expressions for $A,$ $B,$ $C,$ and $D$. 
 
 \medskip
 
The state variables $x(t)$ of a physical realizable system of the form (\ref{eqn:plant}) should satisfy 
the following equal-time commutation relations described by the real antisymmetric matrix 
$\Theta$
\begin{equation}~\label{eq:eqc}
		\left[ x_i(t),x_j(t) \right] 
		= x_i(t) x_j(t) - x_j(t) x_i(t) 
		= 2i\Theta_{ij},\quad\forall t\geq 0
\end{equation}
where $\Theta$ can be of the two following forms: 
\begin{itemize}
	\item [(i)] Canonical, if $\Theta={\rm{diag}}(J,\cdots, J),$ which is a  block diagonal matrix with each diagonal block	equal to 
		$ J = \left[ \begin{array}{cc} 0 & 1 \\ -1 & 0 \end{array} \right]$, or
\item [(ii)] Degenerate canonical, if 
	$\Theta ={\rm{diag}}(0_{n'\times n'},\cdots, J),$ with $0< n'\leq n_x$.
\end{itemize} 
Also, we have the following It\={o} table for $dw$: 
$$\dif w(t) \dif w(t)^T = F_w \dif t.$$ 
Here $F_w$ is a non-negative hermitian matrix and
$F_w = S_w + i T_w,$ where $S_w$ and $T_w$ are the real and imaginary parts of 
$F_w$. The commutation relations for $\dif w(t)$ 
are determined by $T_w$ and the intensity of the noise processes is described 
by $S_w$. We will consider the case where the inputs are thermal noises with 
$S_w,$ a block diagonal matrix with each diagonal block equal to 
$$\left[ \begin{array}{cc}
		1 + 2 k_n & 0 \\ 0 & 1 + 2 k_n
\end{array} \right].$$
The matrix $S_w$ was derived by using the quantum It\={o} multiplication table given in~(\ref{eq:table}).

\medskip

We set the following conventions: 
\begin{itemize} 
\item [(i)] the dimensions $n_x$, $n_y$ and $n_w$ are even; and 
\item [(ii)] $n_y \le n_w$. 
\end{itemize}
Also, without loss of generality, we restrict our attention to quantum plants (\ref{eqn:plant}) with 
canonical commutation relations. This just fixes the choice of basis for $x(t)$. For systems with 
degenerate canonical $\Theta,$ there exists an equivalent description (\ref{eqn:plant}) with 
canonical~$\Theta$ which can be obtained by applying the appropriate state transformation 
(see~\cite{JNP1} for more details).

%\medskip

\section{Physical realizability}~\label{sec:pr}
Not all QSDEs of the form (\ref{eqn:plant}) represent 
the dynamics of physically meaningful open quantum systems.  
As in \cite{JNP1,VuP12c}, 
QSDEs that describe open quantum harmonic oscillators are said to be \emph{physically realizable}. 
Physical Realizability is equivalent to the condition that Equation (\ref{eqn:plant}) 
can be derived from a unitary adapted quantum stochastic evolution 
as described in~(\ref{eq:evox}) and~(\ref{eq:evoy}). 
We restrict our attention to quantum plants which are physically realizable and as such the QSDEs 
(\ref{eqn:plant}) are assumed to be physically realizable. 

In some of the literature (e.g. \cite{JNP1}), the term 
\emph{physical realizability} is used to describe physically meaningful systems with both classical 
and quantum degrees of freedom.
In the following, we give the definition of 
\emph{physical realizability} for QSDEs representing fully quantum systems. 
%Furthermore, we tailor our definition to the class of systems presented by Equation~(\ref{eqn:plant}).
For a more general definition of \emph{physical realizability} see~\cite [Definition~3.1]{JNP1}.

\begin{definition}[\cite{JNP1}] \label{def:pr}
\rm The system described by (\ref{eqn:plant}) is \emph{physically realizable} if 
	it has canonical commutation relations and it represents an open quantum harmonic oscillator.  The system~(\ref{eqn:plant}) describes an open quantum harmonic oscillator if there exists 
	%a non-singular, skew symmetric commutation matrix $\Theta$, 
	a quadratic Hamiltonian $H = \frac{1}{2} x(0)^T R x(0)$,
	with a real, symmetric, $n_x \times n_x$ matrix $R,$ and a  
		coupling operator $L = \Lambda x(0)$, 
		with a complex-valued $\frac{1}{2} n_w \times n_x$ 
	coupling matrix $\Lambda$ 
%	, and a scattering operator $S$ where $S$ is a unitary matrix 
	such that
	\begin{eqnarray}
	x_k(t) &=& U(t)^\dagger (x_k(0) \otimes 1) U(t),\quad k=1,\cdots,n_x \nonumber \\
	y_l(t) &=& U(t)^\dagger (1 \otimes w_l(t)) U(t), \quad l=1,\cdots,n_y \label{eqn:unitary}
\end{eqnarray}
where $\{U(t),\quad t\geq 0\}$ is an adapted process of unitary operators satisfying the following QSDE~\cite{HP84}
	\begin{equation*}
	dU(t)=\left(-iH\,dt-\frac{1}{2}L^\dag L\,dt+[-L^\dag\,\, L^T]\Gamma dw(t)\right)U(t),\quad U(0)=I.
	\end{equation*}
In this case, the matrices $A$, $B$, $C$ and $D$ are given by
	\begin{eqnarray*}
		A = 2 \Theta \left(R + \mathfrak{Im}\left(
			\Lambda^{\dagger}\Lambda \right) \right),
			\label{eqn:a2} \\
		B = 2i \Theta \left[ \begin{array}{cc}
			-\Lambda^{\dagger} & \Lambda^T \end{array} \right] \Gamma,
			\label{eqn:b2} \\
		C = P^T \left[ \begin{array}{cc}
				\Sigma & 0 \\ 0 & \Sigma \end{array} \right]
			\left[ \begin{array}{cc} \Lambda + \Lambda^\# \\
				-i\Lambda + i\Lambda^\# \end{array} \right] ,
			\label{eqn:c2} \\
		D = \left[\begin{array}{cc}
				I_{n_y \times n_y} &
				0_{n_y \times (n_w - n_y)}
			\end{array} \right].
	\end{eqnarray*}
	Here, 
	$\Gamma$ is a 
	$n_w \times n_w$ matrix and 
	\begin{eqnarray*}
	\Gamma &=&  P \diag (M), \\ 
	M &=&  \frac{1}{2}
		\left[ \begin{array}{cc} 1 & i \\ 1 & -i \end{array} \right], \\
	\Sigma &=&  \left[ \begin{array}{cc}
				I_{\frac{1}{2}n_y \times 
				\frac{1}{2}n_y} &
				0_{\frac{1}{2}n_y \times \frac{1}{2}\left(
			n_w - n_y \right) } \end{array} \right].
	\end{eqnarray*}
\rm $P$ is the appropriately dimensioned square permutation 
	matrix such that 
	$$ P \left[ \begin{array}{cccc}a_1 & a_2 & \cdots & a_{2m} \end{array} \right] 
		= \left[ \begin{array}{cccccccc} a_1 & a_3 & \cdots & a_{2m-1} &
		a_2 & a_4 & \cdots & a_{2m} \end{array} \right]$$
	and 
	$\diag (M)$ is the appropriately dimensioned square block diagonal 
	matrix with the matrix $M$ occurring along the diagonal. (Note:  
	dimensions of $P$ and $\diag (M)$ can always be determined from the
	context in which they appear.) Also, $\mathfrak{Im}\left(.\right)$ 
	denotes the imaginary part of a matrix, 
	${X}^\#$ denotes the complex conjugate of a matrix $X$, 
	and ${X}^\dagger$ denotes the 
	complex conjugate transpose of a matrix $X$.
\end{definition}
Note that the equal-time canonical commutation relations~(\ref{eq:eqc}) can be derived from the above definition.

\medskip

For clarity's sake, from this point onward, 
we will often omit time dependence in our notation. We will 
use $x$ in place of $x(t),$ etc. However, the reader should bear in mind that 
in general all operators ($x,y,\dif w$,  etc.) are time dependent. 
Matrices ($A,B,$ etc.) are time invariant.
\section{Problem formulation}~\label{sec:formulation}
In this section, we first define a coherent quantum observer and introduce a class of coherent quantum 
observers that we consider in this paper (see also \cite{Miao:2012ug,Vladimirov:2013ti}). We then present 
the design approaches that we will make use of them in the following section.

\subsection{Coherent quantum observers}
\rm Denote the initial state of the coherent quantum observer by $\xi(0)$ which satisfies the canonical commutation relation, i.e., 
$$\xi(0)\xi(0)^T-(\xi(0)\xi(0)^T))^T=2i\Theta.$$
Also, take the notation $\langle X \rangle_\rho=\tr(\rho X)$ 
corresponding to the quantum expectation 
of an observable $X$ over the density matrix $\rho$ for the
initial joint plant and observer states (see e.g., ~\cite{sakurai1985modern,JNP1}). Then, a coherent quantum observer is a quantum system with an internal state $\xi$ that is designed to estimate the internal variable of the plant's dynamics~(\ref{eqn:plant}) with the following properties:
 \begin{itemize}
	\item [(i)] it is designed such that the tracking error estimation $\langle x - \xi \rangle_\rho$ 
		of the plant dynamics~(\ref{eqn:plant}) exponentially converges to zero in the sense of 
		expected values;%
	\item [(ii)] %
		%Removed for final version: {\color{red} it has second order dynamics} such that 
		the following limit exists, 	
	\begin{equation}
	\bar{J} = \lim_{t \to \infty} 
		\frac{1}{2} 
		\left[
			\left\langle 
				\left( x - \xi \right) 
				\left( x - \xi \right)^T 
			\right\rangle_\rho + 
			\left\langle 
				\Big(\left( x - \xi \right)
				\left( x - \xi \right)^T\Big)^T 
			\right\rangle_\rho
		\right].
	\label{eqn:metric}
	\end{equation}
	Here $\bar{J}$ is a performance metric which  
corresponds to the steady-state quantum expectation of the symmetrized error covariance matrix; 
	\item [(iii)] it is physically realizable.
\end{itemize}

In this paper, we consider the class of coherent quantum observers described by QSDEs of the following form,
\begin{eqnarray}
	\dif \xi &=& \hat{A} \,\xi \dif t 
			+ \hat{B} \dif y 
			+ B_{v_1} \dif v_1
			+ B_{v_2} \dif v_2, \nonumber \\
	\dif \eta &=& \hat{C} \xi \dif t + \dif v_1,
	\label{eqn:observer}
\end{eqnarray}
%(These QSDEs represent an important class of quantum systems considered in \cite{JNP1,NJP1,VuP11a,VuP12a,VuP12c}.)
which are a special case of the QSDEs (\ref{eqn:plant}).
Here $\xi$,$\dif \eta$,$\dif v_1$ and $\dif v_2$ 
are column vectors with dimensions 
$n_\xi$, $n_\eta$, $n_{v_1}$ and $n_{v_2}$ respectively, where
$n_\xi$, $n_\eta$, $n_{v_1}$ and $n_{v_2}$ are even.
Also, we assume $n_\xi = n_x $ and $n_{v_1}=n_\eta$. 
The matrices $\hat{A}$, $\hat{B}$, $B_{v_1}$, $B_{v_2}$ and $\hat{C}$ are real.
The vectors $\xi$,$\dif \eta$,$\dif v_1,$ and $\dif v_2$ each consist of entries which 
are self-adjoint operators acting on the tensor product Hilbert Space 
$\mathcal{H} \otimes \mathcal{F} \otimes \mathcal{H}_2 \otimes \mathcal{F}_2$.
The complex Hilbert space $\mathcal{H}_2$ is the initial space of the coherent quantum observer $\xi(0)$ and 
$\mathcal{F}_2$ is the Boson Fock space which corresponds to the fields other than the output of the plant, 
which also interact with coherent quantum observer. The vectors $\dif y$,  $\dif v_1$ and $\dif v_2$ represent the input fields which interact with the 
coherent quantum observer.
The vector $\dif y$ corresponds to the output of the plant~(\ref{eqn:plant}). 
The entries of $\dif v_1$ and $\dif v_2$ correspond to the quadratures of the 
annihilation and creation processes which act on the boson Fock Space $\mathcal{F}_2$ which are supposed initially in the vacuum 
states. As such, $\dif v_1$ and $\dif v_2$ correspond to quantum vacuum noises.
%The coherent quantum observers that we have proposed in above, also have additional inputs driven by quantum vacuum noises. 
For notational convenience, we separate the vacuum noises into the two vectors 
$\dif v_1$ and $\dif v_2$ such that $n_{v_1} = n_\eta\quad {\rm {and}}\quad n_{v_2} \geq 0$.
These quantum vacuum noises satisfy the It\={o} relations  
$$\dif v_k \dif {v_k}^T = F_{v_k} \dif t,\quad{\rm for}\quad k=1,2,$$ 
where $F_{v_k}$ is a block diagonal matrix with each block equal to 
$$ \left[ \begin{array}{cc} 1 & i \\ -i & 1 \end{array} \right].$$ 
As was the case for the plant, without loss of generality, we restrict our attention to coherent quantum observers 
with canonical commutation relations.

\medskip

The coherent quantum observer may incorporate additional inputs (other than those connected to the plant outputs) 
driven by quantum vacuum noises. These may be required to ensure \emph{physical realizability}. Note that the convergence of the coherent quantum observer is independent of any additional quantum noises in the observer. However, these quantum noises can have an important effect in the value of the performance defined in~(\ref{eqn:metric}).

\subsection{Approaches to design coherent quantum observers}
%In this paper, we consider the problem of designing a coherent quantum observer where the output of the observer 
%provides an estimate of the plant's state. 
Finding an optimal estimation of the plant's state is difficult because of 
the requirement for \emph{physical realizability} and the constraints that this imposes. 
We restrict our attention to design of coherent quantum observers of the form (\ref{eqn:observer}), which provide 
suboptimal solutions to such an estimation.

%Consider  the plant~(\ref{eqn:plant}), we design a  
%coherent quantum observer~(\ref{eqn:observer}), 
%such that the signal part of its output $\zeta = \hat{C} \xi$ tracks the internal state of the plant~$x$. 
%The performance of such coherent observer is determined by the performance metric defined in~(\ref{eqn:metric}) by replacing $\xi$ by $\zeta$. Note that if $\widehat C=I,$ then $x-\zeta$ corresponds to the error estimation of the plant observable $x$ by $\xi,$ i.e., $x-\xi.$ In this paper, we choose $\widehat C=I$ for simplicity. The results remain valid for the case $\widehat C\neq I.$

In the following, we will make use of the following results to make the coherent quantum observers proposed in Equation~(\ref{eqn:observer}) physically realizable.
\begin{theorem}~\label{thm:main}(See~\cite[Theorem $3$]{VuP12c})
	\rm Consider an LTI system of the form (\ref{eqn:observer}) where 
	$\hat{A},\hat{B},\hat{C}$ are given. Then, there exists $B_{v_1}$ and $B_{v_2}$ such that the system is physically 
	realizable with canonical commutation matrix $\Theta$, and with 
	$n_{v_2} = r,$ where $r$ 
	is the rank of the matrix 
	$\left( \Theta \hat{B} \Theta \hat{B}^T \Theta - \Theta \hat{A} - \hat{A}^T \Theta
	- \hat{C}^T \Theta \hat{C} \right)$.
	Conversely, suppose that there exists $B_{v_1}$ and $B_{v_2}$ such that the system 
	(\ref{eqn:observer}) is physically 
	realizable with canonical commutation matrix $\Theta$.
	Then $n_{v_2} \ge r,$ where $r$ is the rank of the matrix 
	$\left( \Theta \hat{B} \Theta \hat{B}^T \Theta - \Theta \hat{A} - \hat{A}^T \Theta
	- \hat{C}^T \Theta \hat{C} \right)$.
	This means that it is not possible to choose 
	$B_{v_1}$ and $B_{v_2}$ such that the system is physically realizable and 
	the dimension of $\dif v_2$ is 
	less than $r$.
\end{theorem}
In~\cite{VuP12c}, during the proof of this theorem, we give a method for constructing  
$B_{v_1}$ and $B_{v_2}.$ We described such a method in \ref{app:A}. Again, this method results in the smallest possible 
dimension for $\dif v_2$ (for a given $\hat{A}$,$\hat{B}$,$\hat{C}$) 
such that (\ref{eqn:observer}) is physically realizable.

\medskip

Below, we give another theorem that we will need in the following.
\begin{theorem}~\label{thm:tf}(See \cite[Theorem $2$]{VuP11a})
	\rm Consider an LTI system of the form (\ref{eqn:observer}), where 
	$\hat{A},\hat{B},\hat{C}$ are given and the system commutation matrix $\Theta$ is
	canonical. 
	Suppose that the Riccati equation 
	\begin{equation}
		X \hat{B} \Theta \hat{B}^T X 
		- \hat{A}^T X - X \hat{A} - \hat{C}^T \Theta \hat{C} = 0 
		\label{eqn:ric1}
	\end{equation}
	has a solution $X$ which is skew-symmetric and suppose that there exists 
	a real non-singular matrix $T$ such that $X = T^T \Theta T$.
	Then, there exists a system described by $\left\{ \tilde{A},\tilde{B}, 
	\tilde{C} \right\}$ with the same transfer function as the system 
	$\left\{ \hat{A},\hat{B},\hat{C}\right\}$ which can be physically realized 
	without the $B_{v_2} \dif v_2$ term (i.e. with $n_{v_2} = 0$) and where
	\begin{eqnarray*}
		X &=&  T^T \Theta T, \\
		\tilde{A} &=& T\hat{A}T^{-1}, \quad
		\tilde{B} = T\hat{B}, \quad
		\tilde{C} = \hat{C}T^{-1},\quad\rm{and}\\
		\tilde{B}_{v_1} &=& \Theta \tilde{C}^T \diag (J).
	\end{eqnarray*}
\end{theorem}
In \cite{VuP12a}, sufficient conditions are given for the existence of a suitable solution to (\ref{eqn:ric1}). 
The accompanying proof leads to a numerical process for obtaining the solution $X$ that we include 
in \ref{app:B}. %The process includes three assumptions which, if satisfied, guarantee the existence of 
%a suitable solution $X$ for the Riccati equation (\ref{eqn:ric1}).
%\section{Problem Formulation} \label{sec:formulation}

\medskip

In the following section, we apply the classical Kalman filtering results. This means that we chose $\hat A=A-KC$ and $\hat B=K$ ($K$ corresponds to Kalman gain) such that $A-K C$ be a Hurwitz matrix which ensures that the coherent quantum observer~(\ref{eqn:observer}) tracks exponentially the plant dynamics~(\ref{eqn:plant}) in the sense of expected values. Moreover, the fact that $A-KC$ is Hurwitz, guaranties that the limit defined in~(\ref{eqn:metric}) exists. Also, we choose $\hat C=I.$  
\section{Algorithms to design coherent quantum observers} \label{sec:observers}
In this section, we give three algorithms to design 
coherent quantum observers. 
%The coherent quantum observers are of the form (\ref{eqn:observer}). 
The motivation behind these algorithms is to treat the quantum plants classically to obtain Kalman 
filters and then obtain physically realizable quantum systems by allowing minimal additional 
quantum vacuum noises. Also, thanks to the performance metric defined in~(\ref{eqn:metric}), we are able to compare the error of convergence for these different algorithms.
\subsection*{Algorithm $1$}
This is the simplest algorithm that we present. 
The quantum noises $\dif w$
driving the plant (\ref{eqn:plant}) are treated as classical Wiener processes with intensity 
$S_w = \mathfrak{Re} \left[ F_w \right]$ where $\mathfrak{Re}[.]$ denotes the real part of a 
matrix. We first obtain a standard Kalman filter as follows  (for details of this, see for example \cite{KS72})
\begin{eqnarray}
	\dif \hat x &=& (A-KC) \hat{x}  \dif t 
			+ K \dif y, \nonumber \\
	\dif \hat{y} &=& \hat{x} \dif t,
	\label{eqn:kalman}
\end{eqnarray}
where $K=(Q C^T + V_{12} ) V_2^{-1}.$ Here $Q$ is the solution to the following algebraic Riccati equation,
\begin{equation} \fl
(A - V_{12} V_2^{-1}C) Q 
 + Q (A - V_{12} V_2^{-1}C)^T - Q C^T V_2^{-1} C Q  
+ V_1 - V_{12} V_2^{-1}V_{12}^T=0,
\label{eqn:ric} \end{equation}
where $Q$ is the steady state of the error covariance matrix given as follows
$$ Q = \lim_{t \to \infty}\left\langle (x - \hat{x})
(x - \hat{x})^T \right\rangle_\rho$$
and $$V = \left[ \begin{array}{cc} V_1 & V_{12} \\ {V_{12}}^T & V_2 \end{array} \right] $$ 
describes the intensity of the joint process
$ \left[ \begin{array}{cc}
	B \dif w \\ D \dif w
\end{array} \right]$.
%{\color{red} $dv_1$ should be $dw_2$}
This means that, 
$$
\mathbb{E} \left[ 
\left[ \begin{array} {cc} B \\ D \end{array} \right]
\dif w \dif w^T
\left[ \begin{array} {cc} B \\ D \end{array} \right]^T
\right]
=
\left[ \begin{array} {cc} V_1 & V_{12} \\ V_{12}^T & V_2 \end{array} \right] \dif t.
$$
In particular, 
\begin{eqnarray*}
	V_1 &=&  B S_w B^T,\\
	V_2 &=& D S_w D^T, \quad \mbox{and}\\
	V_{12} &=& B S_w D^T.
\end{eqnarray*}

Note that the Kalman filter~(\ref{eqn:kalman}) is not physically realizable.

\medskip

Now consider the observer~(\ref{eqn:observer}) and replace $\hat A,$ $\hat B,$ and $\hat C$ by the following values
\begin{eqnarray*}
	\hat{A} &=&  A - KC, \\
	\hat{B} &=&  K, \mbox{and} \\
	\hat C &=&  I.
\end{eqnarray*}
We find
\begin{eqnarray}
	\dif \xi &=& (A-KC)\,\xi \dif t 
			+ K \dif y 
			+ B_{v_1} \dif v_1
			+ B_{v_2} \dif v_2, \nonumber \\
	\dif \eta&=& \xi \dif t + \dif v_1.
	\label{eqn:observer1}
\end{eqnarray}
By Theorem \ref{thm:main}, there exists $B_{v_1}$ and $B_{v_2}$ such that the system (\ref{eqn:observer1}) is physically 
realizable. This system is a coherent quantum observer. 
Furthermore, within the class of quantum systems described by the QSDEs 
(\ref{eqn:observer}), this coherent quantum observer has the minimum number of additional quantum noises ($n_{v_1} + n_{v_2}$) 
for our choice of $\left\{\hat{A},\hat{B},\hat{C} \right\}$. 
%By Theorem \ref{thm:main},
%there exists $B_{v_1}$, $B_{v_2}$ such that the system 
%(\ref{eqn:observer}) is physically realizable.
Details for constructing $B_{v_1}$ and $B_{v_2}$ are included in \ref{app:A}.

\medskip

To see that (\ref{eqn:observer1})  is a coherent quantum observer for the plant (\ref{eqn:plant}), 
it only remains to show that 
$\langle x - \xi \rangle_\rho$ converges to zero exponentially.

The combined plant and observer satisfy the following dynamics
\begin{eqnarray}
	\left[ \begin{array}{c}	
		\dif x \\ \dif \xi 
	\end{array} \right] 
	&=& 
	\mathcal{A} 
	\left[ \begin{array}{c}	
		x \\ \xi 
	\end{array} \right] 
	\dif t
	+ \mathcal{B} 
	\left[ \begin{array}{c}	
		\dif w\\ \dif v_1 \\ \dif v_2 
	\end{array} \right] 
	;
	\nonumber \\
	\dif \eta &=& \xi \dif t + \dif v_1;
	\nonumber \\
	\mathcal{A} &=& 
	\left[ \begin{array}{cc}
		A & 0 \\ KC & A-KC
	\end{array} \right] 
	; \nonumber \\
	\mathcal{B} &=& 
	\left[ \begin{array}{ccc}	
		B & 0 & 0 \\
		KD & B_{v_1} & B_{v_2}
	\end{array} \right] . 
	\label{eqn:cl} 
\end{eqnarray}
From (\ref{eqn:cl}), making the necessary substitutions,  $ x - \xi $ satisfies
\begin{eqnarray*}
\dif (x - \xi) &=&   \left( A - KC \right) (x - \xi) \dif t \\
&& {} + 
\left( B - KD \right) \dif w - B_{v_1} \dif v_1 - B_{v_2} \dif v_2.
\end{eqnarray*} 
Hence,
\begin{equation*}
\dif \langle x - \xi \rangle_\rho =  \left( A - KC \right)\langle x - \xi \rangle_\rho \dif t.
\end{equation*}
As a result,
$\langle x - \xi \rangle_\rho$ converges exponentially to zero if $(A-KC)$ is Hurwitz. The fact that  
 $(A - KC)$ is Hurwitz, follows from the properties of the classical Kalman 
filter which was used to choose $K$.

As $A-KC$ is Hurwitz, 
%we have $\lim_{t \to \infty}J=\bar J,$ where 
the limit in (\ref{eqn:metric}) converges, and 
$\bar J$ is the unique symmetric positive definite solution of the following Lyapunov equation
\begin{eqnarray}
	0 &=& \mathcal{A}_e \bar{J} + \bar{J} \mathcal{A}_e^T + 
	\mathcal{B}_e S_{w,v} {\mathcal{B}_e}^T, \nonumber \\
	\mathcal{A}_e &=&  \left( A - KC \right), \nonumber \\
	\mathcal{B}_e &=&  \left[ \begin{array}{ccc} 
			(B - KD) & - B_{v_1} & - B_{v_2} 
	\end{array} \right], \label{eqn:J}
\end{eqnarray}
where 
$$
	\left[ \begin{array}{c}	
		\dif w \\ \dif v_1 \\ \dif v_2 
	\end{array} \right] 
	\left[ \begin{array}{cccc}	
		\dif w^T & \dif v_1^T  & \dif v_2^T 
	\end{array} \right] 
= F_{w,v} \dif t
	\qquad \rm{and} \qquad S_{w,v} = \mathfrak{Re} \left[ F_{w,v} \right].$$  Finally, the system (\ref{eqn:observer1}) 
so obtained is a coherent quantum observer.

\subsection*{Algorithm $2$}
This algorithm is a refinement of the first, introducing a free parameter $\rho$, 
over which we optimize. The purpose of this parameter is to take into account 
the impact of the noise terms $B_{v_1} \dif v_1(t)$ and $B_{v_2} \dif v_2(t)$ 
when designing the Kalman filter. 
These noise terms are equivalent to additional measurement noise in the 
plant (\ref{eqn:plant}), however they cannot be calculated until after 
the Kalman filter is designed and hence are not available to the design process. 

Compared to Algorithm 1, before calculating the Kalman filter, we first 
introduce an additional term into the plant model (\ref{eqn:plant}) to obtain the 
\emph{modified plant}
\begin{eqnarray}
	\dif x &=& A x \dif t + B \dif w, \nonumber \\
	\dif y &=& C x \dif t + D \dif w + 
	\rho \dif \tilde{w}.
	\label{eqn:plant2}
	\end{eqnarray}
Here, $\dif \tilde{w}$ is a vacuum noise source 
with It\={o} product 
$$\dif \tilde{w} \dif \tilde{w}^T = F_{\tilde{w}} \dif t,$$ 
where $F_{\tilde{w}}$ is a block diagonal matrix with each block equal to 
$$ \left[ \begin{array}{cc} 1 & i \\ -i & 1 \end{array} \right].$$ 
Take $S_{\tilde{w}}$ as the real part of $F_{\tilde{w}}$. 
%added
The noise sources $\dif w$ and $\dif \tilde{w}$ are independent.  

In effect, we inflate the value of the plant 
measurement noise when designing the Kalman filter to compensate for the unknown 
noise terms $B_{v_1} \dif v_1(t)$ and $B_{v_2} \dif v_2(t)$.

We now state Algorithm $2$. The following procedure is repeated for different values of 
$\rho > 0$.
\begin{itemize}
\item Obtain the Kalman filter (\ref{eqn:kalman}) for the modified plant (\ref{eqn:plant2}) with $K$ given by
			$$K =(Q C^T + V_{12} )V_2^{-1},$$
	where $Q$ is the solution to the Riccati equation (\ref{eqn:ric}) with 
	\begin{eqnarray*}
			V_1 &=&  B S_w B^T,\\
			V_2 &=& D S_w D^T + \rho^2 I_{2 \times 2},  \quad \mbox{and}\\
			V_{12} &=& B S_w D^T.
		\end{eqnarray*}

\item Obtain $B_{v_1}$ and $B_{v_2}$ as in Algorithm $1$, such that the system 
	\begin{eqnarray}
		\dif \xi &=& (A-KC) \,\xi \dif t 
				+ K \dif y 
				+ B_{v_1} \dif v_1
				+ B_{v_2} \dif v_2, \nonumber \\
		\dif \eta &=& \xi \dif t + \dif v_1
		\label{eqn:observer2}
	\end{eqnarray}
	is physically realizable. This system is a coherent quantum observer.
\item Calculate the performance metric $\bar{J}$ as in Algorithm $1$ by solving 
	the Lyapunov Equation (\ref{eqn:J}). ($\bar{J}$ is calculated for 
	the actual plant (\ref{eqn:plant}) and not for the modified plant 
	(\ref{eqn:plant2})).
\end{itemize}
Finally, we choose the coherent quantum observer (\ref{eqn:observer2}) which gives the least value of $\bar{J}$.

To see that each iteration results in a coherent quantum observer, consider the following: 
from the properties of the classical Kalman filter, $(A - KC)$ remains Hurwitz for $\rho \ge 0$. 
\subsection*{Algorithm $3$}
Our final algorithm attempts to improve performance by reducing the number of 
additional quantum noises incorporated in the coherent quantum observer. Under certain 
sufficient conditions, it is possible to obtain a coherent quantum observer from a 
state transformation of the Kalman filter obtained in Algorithm $1$. This coherent 
observer incorporates the minimum number of additional noises possible 
for a system of the form~(\ref{eqn:observer}): $n_{v_2} = 0$.

Algorithm $3$ proceeds as follows.
\begin{itemize}
	\item Obtain the Kalman filter (\ref{eqn:kalman}) as in Algorithm $1$.
	\item Attempt to find a transformation $T:$ 
		$$\tilde\xi = T \hat{x}, \quad \tilde{A} = T \hat{A} T^{-1}, \quad
		\tilde{B} = T \hat{B}, \quad \tilde{C} = \hat CT^{-1} $$
	such that the system 
	\begin{eqnarray}
		\dif \tilde\xi &=& \tilde{A} \tilde\xi \dif t 
				+ \tilde{B} \dif y + \tilde B_{v_1} \dif v_1, \nonumber \\
				\dif \eta &=& \tilde{C} \tilde\xi \dif t + \dif v_1,
				\label{eqn:observer3}
	\end{eqnarray} 
	is physically realizable for some $\tilde B_{v_1}$. 
	From Theorem~\ref{thm:tf}, if the Riccati equation 
	\begin{equation*}
		X \hat B \Theta \hat B^T X 
		- \hat A^T X - X \hat A - \hat C^T \Theta \hat C = 0 
	\end{equation*}
	has a non-singular, real, skew-symmetric solution $X$, then such a $T$ exists.
	Sufficient conditions and a construction for $T$ are included in \ref{app:B}.
	If the sufficient conditions for $T$ are not satisfied, we revert to Algorithm $1$.
	%\item The system (\ref{eqn:observer3}) is a coherent quantum observer. 
\end{itemize}
 Now take $\xi=\tilde C\tilde \xi.$ Then, we have
 \begin{eqnarray*}
 d\xi&=&\hat A\xi\dif t+\hat B\dif y+\tilde C\tilde B_{v_1}\dif v_1\\
 \dif \eta &=& \xi \dif t + \dif v_1,
 \end{eqnarray*}
 which is equivalent to the following
  \begin{eqnarray}
 d\xi&=&(A-KC)\xi\dif t+K\dif y+T^{-1}\tilde B_{v_1}\dif v_1\nonumber\\
 \dif \eta &=& \xi \dif t + \dif v_1.~\label{eqn:observerm3}
 \end{eqnarray}
%This means that the dynamics given in~(\ref{eqn:observer3}) corresponds to a coherent quantum observer as the signal part of its output track exponentially the internal state of the plant. 
The combined plant, observer dynamics can be described as follows
\begin{eqnarray}
	\left[ \begin{array}{c} 
		\dif x \\ \dif \xi 
	\end{array} \right] 
	&=& 
	\mathcal{A}_2 
		\left[ \begin{array}{c} 
		x \\ \xi
	\end{array} \right] 
	\dif t
	+ \mathcal{B}_2 
	\left[ \begin{array}{c} 
		\dif w \\ \dif v_1 
	\end{array} \right] ,
	\nonumber \\
	\dif \eta &=&  \xi \dif t + \dif v_1,
	\label{eqn:cl2} \nonumber \\
	\mathcal{A}_2 &=& 
		\left[ \begin{array}{cc} 
		A & 0 \\ KC & A-KC 
	\end{array} \right] 
	, \nonumber \\
	\mathcal{B}_2 &=& 
	\left[ \begin{array}{cc} 
		B & 0 \\
		KD & T^{-1}\tilde B_{v_1}
	\end{array} \right]
	. \nonumber
\end{eqnarray}
Now we show that  $\langle x - \xi \rangle_\rho$ 
converges exponentially to zero. Making the appropriate substitutions, we obtain
\begin{eqnarray*}
\dif (x - \xi ) = \left( A - KC \right) (x - \xi ) \dif t 
 +  \left( B - KD \right) \dif w  - T^{-1} \tilde B_{v_1} \dif v.
\end{eqnarray*}
Then, we find
$$
\dif \langle x - \xi \rangle_\rho=  \left( A - KC \right) \langle x - \xi \rangle_\rho \dif t.$$
From the properties of the Kalman filter (\ref{eqn:kalman}), $(A-KC)$ is Hurwitz, 
therefore   
$ \langle x - \xi \rangle_\rho$ converges exponentially to zero for arbitrary initial states and 
(\ref{eqn:observerm3}) is a coherent quantum observer.

We use the same performance metric (\ref{eqn:metric}) as previously.
Once again, the limit in (\ref{eqn:metric}) converges because  $A-KC$ is Hurwitz.  
Finally, $\bar J$ is the unique symmetric positive definite solution to the Lyapunov equation:
\begin{eqnarray*}
	0 &=& \mathcal{A}_e \bar{J} + \bar{J} \mathcal{A}_e^T + 
	\mathcal{B}_e S_{w,v_1} {\mathcal{B}_e}^T, \nonumber \\
	\mathcal{A}_e &=&  \left( A - KC \right), \nonumber \\
	\mathcal{B}_e &=&  \left[ \begin{array}{cccc} 
			(B - KD) & -T^{-1} \tilde B_{v_1} 
	\end{array} \right],
	\nonumber
\end{eqnarray*}
where 
$$
	\left[ \begin{array}{c}	
		\dif w \\ \dif v_1
	\end{array} \right] 
	\left[ \begin{array}{cc}	
		\dif w^T & \dif v_1^T
	\end{array} \right] 
= F_{w,v_1} \dif t
	\qquad \rm{and}\qquad S_{w,v_1} = \mathfrak{Re} \left[ F_{w,v_1} \right].$$
\section{Measurement-based (classical) observer}~\label{sec:classical}
\begin{figure}[h]
		\includegraphics[trim = 0mm 0mm 0mm 0mm, scale=0.8]{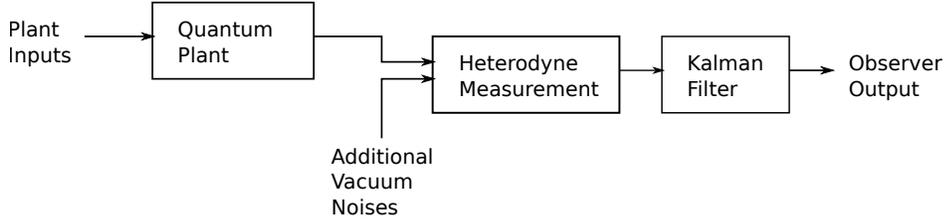}
		\caption{Quantum plant and classical observer consisting of heterodyne measurement 
		and a Kalman filter.}
		\label{fig:classical}	
\end{figure}
In the example which follows, we compare the coherent quantum observers designed in the previous section with 
the following classical observer which consists of heterodyne measurement and a Kalman filter as depicted in 
figure \ref{fig:classical}.

The output of the heterodyne measurement is described by the equation
\begin{equation}
	\dif y_H = \dif y + \dif w_H. \label{eqn:heterodyne}
\end{equation}
where, $\dif w_H$ is a vacuum noise source of dimension 
$n_{w_H} = n_y$ 
and with It\={o} product 
$$\dif w_H \dif w_H^T = F_{w_H} \dif t,$$ 
where $F_{w_H}$ is a block diagonal matrix with each block equal to 
$$ \left[ \begin{array}{cc} 1 & i \\ -i & 1 \end{array} \right],$$
and $S_{w_H}$ is the real part of $F_{w_H}$.

The following Kalman filter is applied to the output~$y_H$ of the heterodyne 
measurement
\begin{eqnarray}
	\dif \hat{x} &=& (A - KC) \hat{x}  \dif t 
			+ K \dif y_H, \nonumber \\
	\dif \hat{y} &=& \hat{x} \dif t.
	\label{eqn:classical_kalman}
\end{eqnarray}
Here, $K=(Q C^T + V_{12} ) V_2^{-1},$ and $Q$ is the solution to the Riccati equation (\ref{eqn:ric}) with 
\begin{eqnarray*}
		V_1 &=&  B S_w B^T, \\
		V_2 &=& D S_w D^T + S_{w_H}, \quad \mbox{and}\\
		V_{12} &=& B S_w D^T.
	\end{eqnarray*}
By combining equations (\ref{eqn:plant}), (\ref{eqn:heterodyne}) and 
	(\ref{eqn:classical_kalman}), we obtain the following dynamics for the combined 
	plant and classical observer
\begin{eqnarray*}
	\left[ \begin{array}{c}	
			\dif x \\ \dif \hat{x} 
	\end{array} \right] 
	&=& 
	\mathcal{A} 
	\left[ \begin{array}{c}	
		x \\ \hat{x} 
	\end{array} \right] 
	\dif t
	+ \mathcal{B} 	
	\left[ \begin{array}{c}	
			\dif w \\ \dif w_H  
	\end{array} \right]
	,
	\nonumber \\
	\mathcal{A} &=& 
	\left[ \begin{array}{cc}
		A & 0 \\ KC &A-KC
	\end{array} \right] 
	, \nonumber \\
	\mathcal{B} &=& 
	\left[ \begin{array}{cc}	
		B & 0 \\
		KD & K
\end{array} \right]
	. 
\end{eqnarray*}
The performance metric 
$$J =  \lim_{t \to \infty} \left\langle (x - \hat{x})(x - \hat{x})^T\right \rangle_\rho,$$
for the classical observer is the solution to the 
following Lyapunov equation:
\begin{eqnarray*}
	0 &=& \mathcal{A}_e J + J \mathcal{A}_e^T + 
	\mathcal{B}_e S_{w,w_H} {\mathcal{B}_e}^T, \nonumber \\
	\mathcal{A}_e &=&  \left( A - KC \right), \nonumber \\
	\mathcal{B}_e &=&  \left[ \begin{array}{cc} 
			B - KD & - K
	\end{array} \right],
	\nonumber
\end{eqnarray*}
where 
$$
	\left[ \begin{array}{c}	
		\dif w \\ \dif w_H
	\end{array} \right] 
	\left[ \begin{array}{ccc}	
		\dif w^T & \dif w_H^T
	\end{array} \right] 
	= F_{w,w_H} \dif t
	\qquad\rm{and}\qquad S_{w,w_H} = \mathfrak{Re} \left[ F_{w,w_H} \right].$$
%%%%%%%%%%%%%%%%%%%%%%%%%%%%%%
\section{Example} \label{sec:example}
\begin{figure}[h]
		\includegraphics[trim = 0mm 0mm 0mm 0mm, scale=0.8]{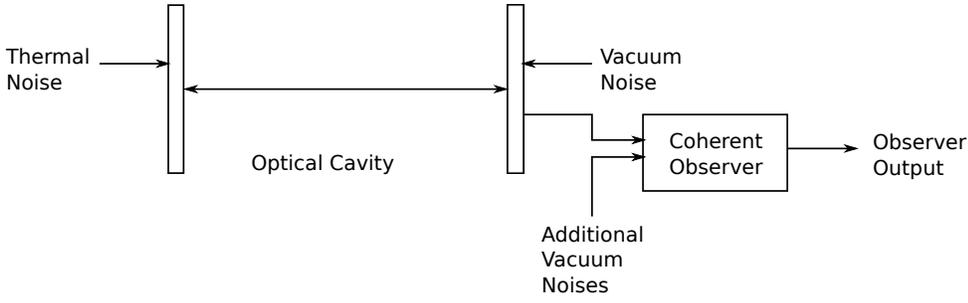}
		\caption{Plant and coherent quantum observer configuration.}
		\label{fig:ex}	
\end{figure}
Consider the quantum plant depicted in Figure \ref{fig:ex}. This plant consists of an optical cavity with thermal and vacuum 
noise inputs.
Its dynamics are described by the following QSDEs of the form (\ref{eqn:plant})
\begin{eqnarray}
	\dif x &=& -\frac{1}{2}\left( \kappa_1 + \kappa_2 \right) x \dif t 
	- \sqrt{\kappa_1} \dif w_1 - \sqrt{\kappa_2} \dif w_2, \nonumber \\
	\dif y &=& \sqrt{\kappa_1} x \dif t + \dif w_1.
	\label{eqn:example_plant}
\end{eqnarray}
Here, $\kappa_1, \kappa_2$ are related to the mirror reflectances, $\dif w_1$ is vacuum noise and 
$\dif w_2$ is thermal noise of intensity $k_n$,
$$S_{w_1} = I_{2 \times 2} \qquad \rm{and} \qquad S_{w_2} = (1 + 2 k_n) I_{2 \times 2}.$$

We consider three scenarios, each with different values for $\kappa_1, \kappa_2$. 
For each scenario, we apply our algorithms to obtain coherent quantum observers 
across a range of thermal noise intensities $k_n$.
\subsection{Scenario 1: $\kappa_1 = \kappa_2 = 0.1$}
	\begin{figure}[h]
		\includegraphics[trim = 0mm 0mm 0mm 0mm, scale=0.6]{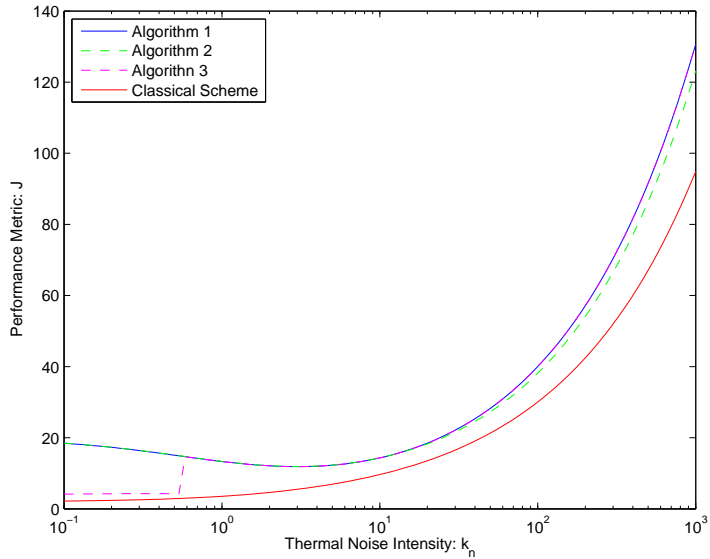}
		\caption{Comparison of observers for $\kappa_1 = \kappa_2 = 0.1.$}
		\label{fig:ex1}
	\end{figure}
Figure \ref{fig:ex1} compares the performance metric $J$ for each of our coherent quantum observers with that for a classical observer 
consisting of heterodyne measurement and a Kalman filter. The classical observer performs best.
This is not surprising as each of our coherent quantum observers introduces at least as much additional quantum noise 
as does the heterodyne measurement in the classical observer. Furthermore, the classical observer is optimal solution with respect to the 
output of the heterodyne measurement whereas the coherent quantum observers we consider are suboptimal.
%
% move the following sentence later??? (or just delete?)
%
Notwithstanding this result, it is still of interest to develop tractable 
methods for designing coherent quantum observers as other considerations may favour the use of 
coherent quantum observers over measurement-based observers, in particular when the controllers are added.

\begin{figure}[h]
	\includegraphics[trim = 0mm 0mm 0mm 0mm, scale=0.6]{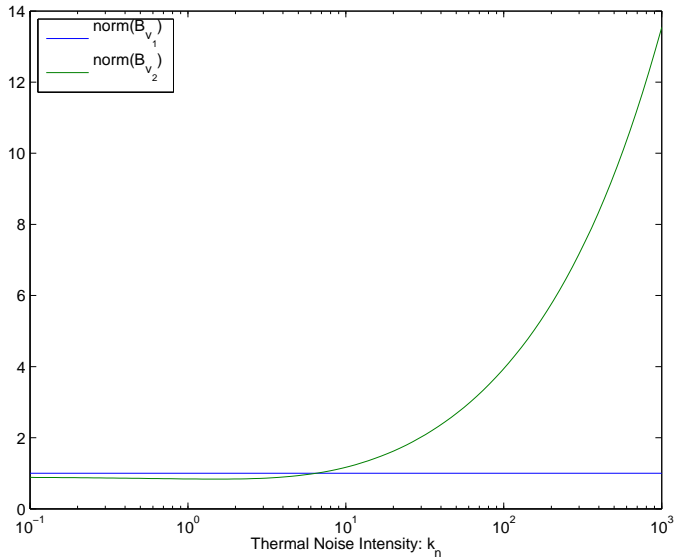}
	\caption{Significance of $B_{v_1}$ and $B_{v_2}$ for different values of $k_n.$}
	\label{fig:ex1b}
\end{figure}

\begin{figure}[h]
	\includegraphics[trim = 0mm 0mm 0mm 0mm, scale=0.6]{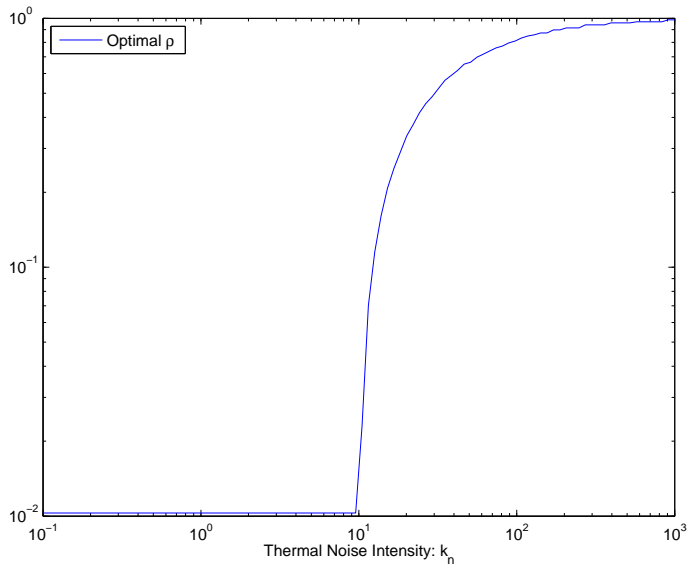}
	\caption{Choice of $\rho$ for different values of $k_n.$}
	\label{fig:ex1c}
\end{figure}

The performance of Algorithm $1$ never exceeds that of Algorithm $2$. This is because, in Algorithm $2$, $\rho = 0$ results in the same 
coherent quantum observer as Algorithm $1$. Recall that Algorithm $1$ does not take into account the $B_{v_1}$ and $B_{v_2}$ terms when 
designing the Kalman filter. Figure \ref{fig:ex1b} shows the matrix norms for $B_{v_1}$ and $B_{v_2}$ for different values of $k_n$. 
It seems reasonable that as $B_{v_2}$ becomes more significant, there is greater scope for Algorithm $2$ to outperform Algorithm $1$. 
This explains why Algorithm $2'$s relative performance increases with $k_n$. 
Figure \ref{fig:ex1c} shows how the optimal $\rho$ varies with $k_n$ in Algorithm $2$.
We now turn our attention to Algorithm $3$. For small values of $k_n$, a suitable transformation matrix $T$ 
 was found and a coherent quantum observer obtained with $n_{v_2} = 0$. In this regime, 
Algorithm $3$ outperforms the other algorithms and produces a coherent quantum observer which approaches the performance 
of the classical observer.
The discontinuity in Algorithm $3'$s performance corresponds to the point above which, no suitable $T$ was found. 
In this regime the algorithm produces the same coherent quantum observer as Algorithm $1.$ 
The range of $k_n$ for which a suitable $T$ exists is dependent on $\kappa_1$ and $\kappa_2$ as demonstrated in the 
scenarios which follow.

Note that Algorithm~3 produces a 
coherent quantum observer with a different value for $B_{v_1}$ than that from Algorithm~1. In the following scenarios we 
shall see that in some regimes, despite introducing a smaller number of vacuum noises, Algorithm~3 does not 
perform better than Algorithm~1.

Finally, we briefly comment on the performance of the observers in the limit as $k_n$ approaches zero (that is, as 
the noise input $\dif w_2$ approaches a vacuum noise). For $k_n = 0$, the Kalman filter 
gain $K$, obtained in our algorithms, is zero. When $\dif w_2$ is a vacuum noise, the output of the plant gives no 
useful information about the internal state of the 
plant. In this special case, the optimal coherent quantum observer is the trivial one: a vacuum noise source. 
See \cite{Pet13a} for a discussion of a class of plants driven solely by vacuum noises for which 
 the authors show that the optimal controllers (and by implication the optimal observers) are trivial ones.
\subsection{Scenario 2: $\kappa_1 = 0.5$; $\kappa_2 = 0.01$}
\begin{figure}[h]
		\includegraphics[trim = 0mm 0mm 0mm 0mm, scale=0.6]{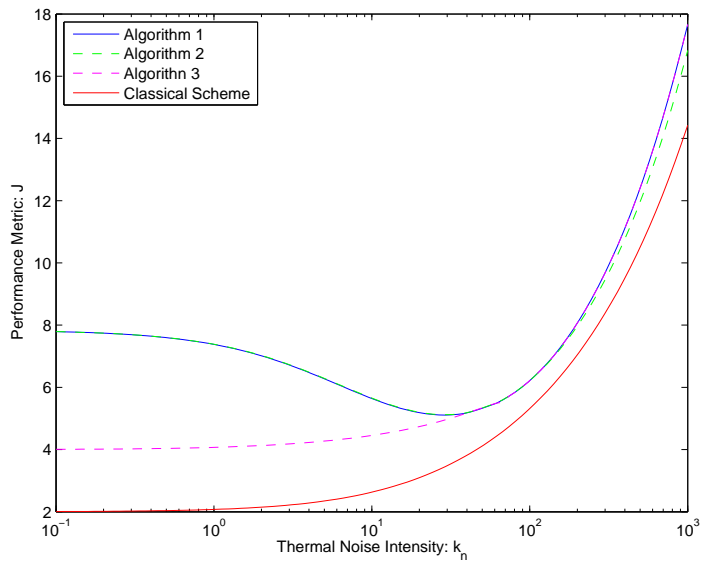}
		\caption{Comparison of observers for $\kappa_1 = 0.5, \kappa_2 = 0.01.$}
		\label{fig:ex2}
	\end{figure}

\begin{figure}[h]
		\includegraphics[trim = 0mm 0mm 0mm 0mm, scale=0.6]{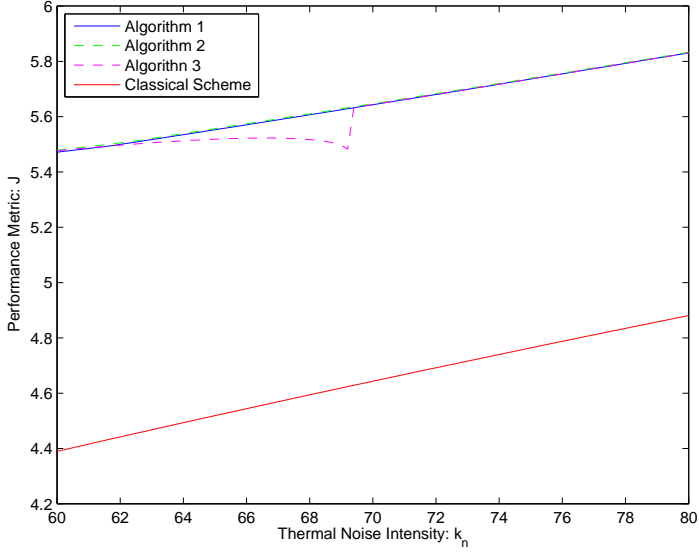}
		\caption{Comparison of observers for $\kappa_1 = 0.5, \kappa_2 = 0.01.$}
		\label{fig:ex2b}
	\end{figure}

Figure \ref{fig:ex2} shows the performance of the observers obtained for $\kappa_1 = 0.5$ and $\kappa_2 = 0.01$. 
(Compared to Scenario 1, mirror 1 is more lossy, while mirror 2 is less lossy.) 
For these mirrors, Algorithm~3 performs better than Algorithm~1 for greater noise intensities $k_n$.  
The discontinuity where no suitable state transformation $T$ was found in Algorithm~3 
occurs at $k_n = 69$ and is shown in more detail in Figure \ref{fig:ex2b}. 
\subsection{Scenario 3: $\kappa_1 = 0.8$; $\kappa_2 = 0.01$}
\begin{figure}[h]
		\includegraphics[trim = 0mm 0mm 0mm 0mm, scale=0.6]{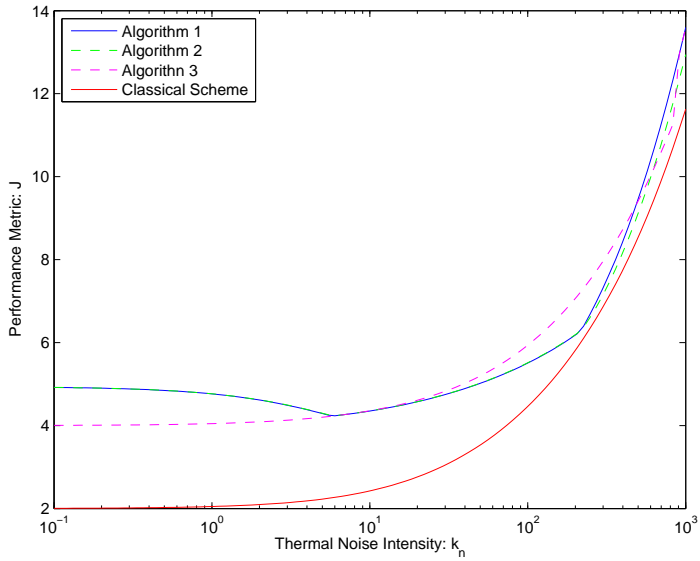}
		\caption{Comparison of observers for $\kappa_1 = 0.8, \kappa_2 = 0.01.$}
		\label{fig:ex3}
	\end{figure}

Figure \ref{fig:ex3} shows the performance of the observers obtained for $\kappa_1 = 0.8$ and $\kappa_2 = 0.01$. 
Compared to the previous scenarios, mirror 1 is even more lossy. As a result, the discontinuity in Algorithm 3's performance, 
above which no suitable state transformation $T$ was found, occurs at the increased noise intensity $k_n = 910$. 
Below this point, Algorithm~3 gives a coherent quantum observer with $n_{v_2} = 0$ while above this point it gives a coherent quantum observer 
with $n_{v_2} = 2$. Algorithms~1 and 2 give coherent quantum observers with $n_{v_2} = 2$ for all considered values of $k_n$.

This scenario demonstrates a region where Algorithm~2 performs better than Algorithm~3 despite the latter 
giving a coherent quantum observer with less quantum noise sources. This is because, in this region, the impact of the $B_{v_1}$ term 
obtained in Algorithm~3 is more significant than the combined impact of both the $B_{v_1}$ and $B_{v_2}$ terms in Algorithm~2. 

Finally, this scenario suggests that the performance metric $J$ obtained for Algorithms~1 and 2 is not necessarily smooth 
with respect to $k_n$. Obtaining an explanation for this observation remains the subject of future research.
\section{Conclusions} \label{sec:conclusion}
Like the celebrated Kalman filter in the context of classical feedback control problems, 
it is envisaged that coherent quantum observers will 
play a pivotal role in solving coherent quantum feedback control problems. Here, we have proposed 
three algorithms for the design of coherent quantum observers. The key idea behind each of our algorithms 
was to first treat the quantum plants classically to obtain a Kalman filter. We then made use of previous 
results to obtain a physically realizable system by taking the Kalman filter obtained and allowing additional 
vacuum noise sources in its quantum implementation. 
Algorithms~2 and 3 incorporate refinements to Algorithm~1 in an attempt to improve 
performance.

We compare the performance of the coherent quantum observers obtained with a measurement-based (classical) observer 
by way of an example involving an optical cavity with thermal and vacuum noise inputs. For each of the scenarios 
considered, the classical observer performs best. Algorithm~2 always performs at least as well as Algorithm~1. 
Algorithm~3 can potentially give a coherent quantum observer with a smaller number of quantum vacuum noise inputs than 
the other algorithms, however this does not guarantee better performance.

\ack
The authors gratefully acknowledge Professors Hideo Mabuchi, Ian Petersen and Hendra Nurdin 
for helpful discussions. Shanon L. Vuglar gratefully acknowledges support by the Australian Research Council 
and the Air Force Office of Scientific Research 
(Grant Nos. FA2386-09-1-4089 and FA2386-12-1-4075). Hadis Amini has a Math$+$X postdoctoral fellowship from the Simons
Foundation.

\appendix
\section{}\label{app:A}
Suppose we have a system of the form (\ref{eqn:observer}) with canonical 
commutation matrix $\Theta$ and where 
$\hat{A},\hat{B},$ and $\hat{C}$ are given. 
The following construction for $B_{v_1}$ and $B_{v_2}$ results in 
a physically realizable system. It is not possible to construct $B_{v_1}$ and 
$B_{v_2}$ with smaller $n_{v_2}$ such that (\ref{eqn:observer}) is physically 
realizable. For further details see \cite{VuP12c}.
\begin{itemize}
\item 
Construct the matrix
$$ \tilde{S} = \Theta \hat{B} \Theta \hat{B}^T \Theta - \Theta \hat{A} - \hat{A}^T \Theta
- \hat{C}^T \Theta \hat{C}.$$
(Here $\Theta$ is the  
canonical commutation matrix of dimension $n_x \times n_x$) 
\item 
	Find the rank of the matrix $\tilde{S}$: $n_{v_2} = \mbox{rank} \left[ \tilde{S} \right]$.
\item
	Calculate $S = \frac{i}{4}\tilde{S}$.
\item
	Diagonalize $S$: $S = U^{\dagger}DU$. Here $D$ is diagonal and $U$ is unitary.
\item	Construct $\hat{D}$ by replacing each element of $D$ with its absolute value. 
\item	Construct  $W = \left( \hat{D} + D \right)^{\frac{1}{2}} U$. 
\item	Construct $B_{v_1}$ and $B_{v_2}$ as follows: 
	\begin{eqnarray*}
		B_{v_1} &=& \Theta \hat{C}^T \diag (J); \\
		B_{v_2} &=& 2i\theta 
		\left[ 
		\begin{array}{cc}
		-W^{\dagger} 
		& 
		W^T 
		\end{array} 
		\right]
			P \nonumber \diag (M).
	\end{eqnarray*}
\end{itemize}
\section{}\label{app:B}
Here we give a numerical process for obtaining a suitable solution $X$ to the Riccati equation~(\ref{eqn:ric1}) in 
Theorem~\ref{thm:tf}. The three assumptions in the following, guarantee the existence of the solution $X$.
For further details see \cite{VuP12a}.
\begin{itemize}
\item
	Construct 
	$$Z = \left[ \begin{array}{cc}\hat{A} & -\hat{B} \Theta \hat{B}^T \\ 
		-\hat{C}^T \Theta \hat{C} & -\hat{A}^T \end{array} \right].$$
\item
	Find the eigenvalues and eigenvectors of $Z$.
\item	Assumption 1: That $Z$ has no purely imaginary eigenvalues.
	In practice, this means 
	checking that the real part of each eigenvalue has magnitude greater than some 
	small numerical tolerance.
\item
	Construct the matrix 
	$$\left[ \begin{array}{c} X_1 \\ X_2 \end{array} \right]$$ 
	such that its 
	columns are the eigenvectors of $Z$ that correspond to eigenvalues with negative 
	real part.
\item
	Assumption 2: That $X_1$ is non-singular.
\item   Calculate $X = X_2 X_1^{-1}$.
\item
	Assumption 3: That $X$ is non-singular.
\item
	Find the eigenvalues and eigenvectors of $X$. Hence, construct diagonal $\Lambda$ 
	with diagonal entries the eigenvalues of $X$ and $V$ with 
	columns the corresponding eigenvectors normalized to length 1. 
\item
	Construct the $n_\xi \times n_\xi$ diagonal matrix $\tilde{\Lambda}$ with alternating 
	diagonal entries $i$ and~$-i$.
\item
	Construct the $n_\xi \times n_\xi$ 
	block diagonal matrix $\tilde{V}$ with 
	each diagonal block corresponding to $\frac{1}{\sqrt{2}}
	\left[ \begin{array}{cc}1 & 1 \\ i & -i
	\end{array} \right]$.
\item
	Calculate $D = \left( \tilde{\Lambda}^{-1} \Lambda \right)^{\frac{1}{2}}$.
\item
	Calculate $T = \tilde{V} D V^{\dagger}$.
\item
	Construct
	\begin{eqnarray*}
		\tilde{A} &=& T \hat{A} T^{-1}, \quad
		\tilde{B} = T \hat{B}, \quad
		\tilde{C} = \hat{C} T^{-1},\quad\rm{and}\\
		\tilde B_{v_1} &=& \Theta \tilde{C}^T \diag (J).
	\end{eqnarray*}
\end{itemize}
The system $\left\{ \tilde{A},\tilde{B},\tilde{C} \right\}$ has the same transfer 
function as $\left\{\hat{A},\hat{B},\hat{C} \right\}$ 
and is \emph{physically realizable} with 
$n_{v_2} = 0$ and with 
$\tilde B_{v_1}$ as constructed above.
\section*{References}
%\begin{thebibliography}{<num>}
%	\item test
%\end{thebibliography}
\bibliography{local}
\bibliographystyle{iopart-num}

\end{document}